\documentclass[preprint,12pt]{elsarticle}
\usepackage{fancyhdr}
\usepackage{fullpage}
\usepackage{amsmath}
\usepackage{amsbsy}
\usepackage{amssymb}
\usepackage{amscd}
\usepackage{amsfonts}
\usepackage{supertabular}
\usepackage{graphics}
\usepackage{verbatim}
\usepackage{epsfig}
\usepackage{xspace}
\usepackage{euscript}
\usepackage{alltt}
\usepackage{boxedminipage}
\usepackage{float}
\usepackage[colorlinks]{hyperref}
\usepackage{color}
\usepackage[all]{xy}
\usepackage{t1enc}
\usepackage{times,exscale}
\usepackage{graphicx,calc}
\usepackage{subfig}
\usepackage[ruled,vlined]{algorithm2e}
\usepackage{epstopdf}
\usepackage{array,multirow}
\usepackage{braket}

\journal{Elsevier}

\begin{document}
%%%%%%%%%%%%%%%%%%%%%%%%%%%%%%%%%%%%%%%%%%%%%%%%%%%%%%%%%%%%%%%%%%%%%%%%%%%%%%%%%%%%%%%%%%%%%%%%%%%%%%%%%%%%%%%%%%%%%%%%%%%%%%%%%5
\begin{frontmatter}
\title{On preconditioning the self-consistent field iteration in real-space Density Functional Theory}
\author{Shashikant Kumar}
\author{Qimen Xu}
\author{Phanish Suryanarayana\corref{cor}}

\address[gatech]{College of Engineering, Georgia Institute of Technology, Atlanta, GA 30332, USA}
% \address[llnl]{Physics Division, Lawrence Livermore National Laboratory, Livermore, CA 94550, USA}
\cortext[cor]{Corresponding Author (\it phanish.suryanarayana@ce.gatech.edu) }

\begin{abstract}
We present a real-space formulation for isotropic Fourier-space preconditioners used to accelerate the self-consistent field iteration in Density Functional Theory calculations. Specifically, after approximating the preconditioner in Fourier space using a rational function, we express its real-space application in terms of the solution of sparse Helmholtz-type systems.  Using the truncated-Kerker and Resta preconditioners as representative examples, we show that the proposed real-space method is both accurate and efficient, requiring the solution of a single linear system, while accelerating self-consistency to the same extent as its exact Fourier-space counterpart. 
\end{abstract}

        
\end{frontmatter}
%%%%%%%%%%%%%%%%%%%%%%%%%%%%%%%%%%%%%%%%%%%%%%%%%%%%%%%%%%%%%%%%%%%%%%%%%%%%%%%%%%%%%%%%%%%%%%%%%%%%%%%%%%%%%%%%%%%%%%%%%%%%%%%%%%%%%%%%

%%%%%%%%%%%%%%%%%%%%%%%%%%%%%%%%%%%%%%%%%%%%%%%%%%%%%%%%%%%%%%%%%%%%%%%%%%%%%%%%%%%%%%%%%%%%%%%%%%%%%%%
\section{Introduction}
Kohn-Sham Density Functional Theory (DFT) \cite{Hohenberg,Kohn1965} is one of the most widely used methods for understanding/predicting a range of material properties and chemical phenomena from first principles. Though DFT has a highly favorable accuracy-to-cost ratio compared to other such ab initio methods, the length-and time-scales that can be reached is still limited by the large computational effort involved. Since the cost of DFT calculations is directly proportional to the number of iterations required by the self-consistent field (SCF) method \cite{Martin2004}---standard fixed-point iteration technique for determining the electronic ground state---there is great incentive to reduce them to as few as possible.  

Extrapolation techniques are typically used to accelerate the convergence of the SCF iteration in DFT, referred to as mixing schemes in this context \cite{anderson1965iterative,broyden1965class,pulay1980convergence,bendt1982new,srivastava1984broyden,vanderbilt1984total,eyert1996comparative,marks2008robust,fang2009two,pratapa2015restarted,banerjee2015periodic}. Though these schemes have been found to be extremely effective and represent significant advances, a number of challenges remain. Specifically, convergence can be slow to non-existent due to charge sloshing effects, particularly as the size of the system is increased for metallic as well as inhomogeneous systems \cite{annett1995efficiency,VASP,lin2013elliptic,Woods_2019}. In order to mitigate this behavior, there have been a number of efforts to develop preconditioners that can be employed in conjunction with mixing schemes to further accelerate the convergence of the underlying fixed-point iteration \cite{lin2013elliptic,kerker1981efficient,ho1982dielectric,vanderbilt1984total,raczkowski2001thomas,sawamura2004second,anglade2008preconditioning,zhou2018applicability}. 

An ideal preconditioner for minimizing the number of SCF iterations is the inverse of the dielectric matrix, a strategy adopted in the HIJ method \cite{ho1982dielectric}. However, this approach has a quartic scaling with system size and is associated with a extremely large prefactor. Indeed, this prefactor can be substantially reduced using the extrapolar method \cite{anglade2008preconditioning}, however the quartic scaling still represents a serious bottleneck. In view of this, there have been efforts directed towards the development of simpler models of the dielectric matrix, resulting in preconditioners that are efficient to compute and apply \cite{kerker1981efficient,raczkowski2001thomas,lin2013elliptic,zhou2018applicability}.
Among these, isotropic Fourier-space preconditioners are perhaps the most popular in Kohn-Sham DFT, with origins in the work of Kerker \cite{kerker1981efficient}, who employed a model dielectric function suitable for simple metallic systems. 

Given the restriction of the Kerker preconditioner to homogeneous electron gas systems, variants such as truncated-Kerker \cite{VASP,zhou2018applicability} and Resta \cite{resta1977thomas,zhou2018applicability} have been developed for preconditioning insulating as well as inhomogeneous systems.\footnote{The Elliptic preconditioner \cite{lin2013elliptic} also has a similar motivation, but is not considered in this work since it already has an efficient real-space formulation.} In fact, the truncated-Kerker preconditioner is currently the default in the established planewave code VASP \cite{VASP}. Since these preconditioners have a diagonal representation in Fourier space, they are easy and efficient to apply within the planewave method, but are unsuitable for real-space methods, where they take a global form. Given that real-space codes are now able to outperform their planewave counterparts by being able to leverage large-scale computational resources \cite{ghosh2017sparc1,ghosh2017sparc2}, while being amenable to the development of linear-scaling methods \cite{osei2014accurate,suryanarayana2018sqdft} and offering the flexibility in choice of boundary conditions \cite{natan2008real,Phanish2012,ghosh2019symmetry}, efficient real-space analogues for such preconditioners are highly desired.

In this work, we overcome the above limitation of real-space methods and develop a real-space formulation for isotropic Fourier-space preconditioners used to accelerate the SCF iteration in DFT calculations.\footnote{A real-space formulation for the original Kerker preconditioner has been developed previously in Ref.~\cite{shiihara2008real}. } In this approach, after approximating the preconditioner in Fourier space using a rational function, we express its real-space application in terms of the solution of sparse Helmholtz-type linear systems of equations. Using the truncated-Kerker and Resta preconditioners as representative examples, we demonstrate that the proposed method is both accurate and efficient. In particular, we show that the application of these preconditioners in real-space requires the solution of only one linear system, while accelerating self-consistency to the same degree as the corresponding exact Fourier-space preconditioner.  

The remainder of the this paper is organized as follows. In Section~\ref{Sec:ONDFT}, we describe the proposed formulation for the real-space application of isotropic Fourier-space preconditioners.  Next, we verify its accuracy and efficiency in Section~\ref{Sec:Results}. Finally, we provide concluding remarks in Section~\ref{Sec:ConcludingRemarks}.

%%%%%%%%%%%%%%%%%%%%%%%%%%%%%%%%%%%%%%%%%%%%%%%%%%%%%%%%%%%%%%%%%%%%%%%%%%%%%%%%%%%%%%%%%%%%%%%%%%%%%%%
\section{Formulation}\label{Sec:ONDFT}
The electronic ground state in Kohn-Sham DFT is typically calculated using the  self-consistent field (SCF) method \cite{Martin2004}, a technique in which the underlying equations are cast as a nonlinear fixed-point problem:
\begin{equation} \label{Eqn:mapping}
\boldsymbol{\rho} = \mathbf{g} ( \boldsymbol{\rho}) \,,
\end{equation}
where $\boldsymbol{\rho} \in \mathbb{R}^{\rm N \times 1}$ is the electron density\footnote{A similar fixed-point problem can be written in terms of the potential.}, and $\mathbf{g}:\mathbb{R}^{\rm N \times 1} \rightarrow \mathbb{R}^{\rm N \times 1} $ is a mapping that comprises of two components: computation of the potential from the electron density and then the computation of the electron density from a Schr\"{o}dinger-type eigenproblem. The solution of Eqn.~\ref{Eqn:mapping} is typically achieved using preconditioned mixing schemes, i.e., an iteration of the form:
\begin{equation} \label{Eqn:mixing-equation}
\boldsymbol{\rho}_{k+1}= \langle\boldsymbol{\rho}_{k}\rangle + \beta\mathbf{P}\langle\mathbf{f}_k\rangle \,,
\end{equation}
where $k$ is the iteration number, $\mathbf{f}_k = \mathbf{g} ( \boldsymbol{\rho}_k) -\boldsymbol{\rho}_k \in \mathbb{R}^{\rm N \times 1}$ is the residual, $\beta \in \mathbb{R}$ is the mixing parameter, $\mathbf{P} \in \mathbb{R}^{\rm N\times N}$ is the preconditioner, and $\langle . \rangle$ denotes some linear combination of the enclosed quantity from previous iterations. Indeed, the coefficients used for the linear combination vary between iterations and are also dependent on the choice of mixing scheme. 

The convergence behavior of the fixed-point problem in Eqn.~\ref{Eqn:mapping} is dictated by the properties of the dielectric matrix $\mathbf{g}'( \boldsymbol{\rho}) \in \mathbb{R}^{\rm N\times N}$ \cite{VASP,lin2013elliptic}. Indeed, the goal of the preconditioner in combination with the mixing scheme is to try and approximate the inverse of this matrix, thereby reducing the SCF iterations required for convergence. In particular, the preconditioner tries to mitigate charge sloshing effects and render the number of iterations independent of system size. Many of the commonly used preconditioners in Kohn-Sham DFT (e.g., Kerker \cite{kerker1981efficient} and truncated-Kerker \cite{VASP,zhou2018applicability}) are isotropic and have a diagonal representation in Fourier space, which makes them easy and efficient to apply in plane-wave methods and codes. However, their equivalent real-space representations are global in nature, i.e., the preconditioner matrices are dense, which makes them impractical for real-space calculations. In view of this, we now develop a general efficient approach for the application of such preconditioners in real-space DFT.

Consider a diagonal preconditioner matrix $\mathbf{\Tilde{P}} \in \mathbb{R}^{\rm N\times N}$ in the Fourier space representation, with entries for a given wavevector $\mathbf{q}$ denoted by the isotropic function $\Tilde{P}(|\mathbf{q}|)$. First, we approximate $\Tilde{P}(|\mathbf{q}|)$ using a rational function of the form:
\begin{align} 
  \Tilde{P}(\mathbf{|q|}) & \approx   \frac{ b_t |\mathbf{q}|^{2t}+b_{t-1} |\mathbf{q}|^{2t-2}+\dotsm + b_1 |\mathbf{q}|^2 + b_0}{|\mathbf{q}|^{2t}+a_{t-1} |\mathbf{q}|^{2t-2}+\dotsm + a_1 |\mathbf{q}|^2 + a_0}  \,, \label{Eqn:Preconditioner-rational} 
\end{align}
where the coefficients $a_j \in \mathbb{R}$ and $b_j \in \mathbb{R}$ are chosen so as to provide the best fit for $\Tilde{P}(|\mathbf{q}|)$. Next, we expand this rational function in terms of partial fractions: 
\begin{equation} \label{Eqn:Preconditioner-partial}
    \Tilde{P}(|\mathbf{q}|) \approx c_0 + \sum_{j=1}^{t} \frac{c_j |\mathbf{q}|^2}{|\mathbf{q}|^2 + d_j}  \,,
 \end{equation}
where the coefficients $c_j \in \mathbb{C}$ and $d_j \in \mathbb{C}$ can be determined in terms of $a_j$ and $b_j$. Noting that the eigenvalues of the negative periodic  Laplacian are $\mathbf{q}^2$,  the real-space representation for Eqn.~\ref{Eqn:Preconditioner-partial} takes the form: 
\begin{equation} \label{Eqn:Preconditioner-real-space}
\mathbf{P} = c_0\mathbf{I} + \sum_{j=1}^{t} c_j(-\mathbf{\nabla}_h^2 + d_j\mathbf{I})^{-1} (-\mathbf{\nabla}_h^2) \,,
\end{equation}
where $\mathbf{\nabla}_h^2 \in \mathbb{R}^{\rm N\times N} $ is the real-space Laplacian and $\mathbf{I} \in \mathbb{R}^{\rm N\times N}$ is the identity matrix. As written in Eqn.~\ref{Eqn:Preconditioner-real-space}, the real-space preconditioner matrix $\mathbf{P}$ is dense, which makes its explicit calculation and application impractical. However, it is clear from Eqn.~\ref{Eqn:mixing-equation} that $\mathbf{P}$ does not need to be calculated, but rather only its product with a vector needs to be determined. In view of this, we arrive at
\begin{equation} \label{Eqn:PreconditionerApplication}
\mathbf{P} \langle\mathbf{f}_k\rangle = c_0 \langle\mathbf{f}_k\rangle + \sum_{j=1}^{t} c_j \langle \mathbf{h}_{k,j} \rangle \,,
\end{equation}
where $\langle \mathbf{h}_{k,j} \rangle $ is the solution of the following sparse Helmholtz-type linear system:
 \begin{equation} \label{Eqn:Preconditioner-Helmholtz}
(-\mathbf{\nabla}_h^2 + d_j\mathbf{I}) \langle \mathbf{h}_{k,j} \rangle = - \mathbf{\nabla}_h^2 \langle \mathbf{f}_k  \rangle\,.
\end{equation}

The application of the isotropic Fourier-space preconditioner in real-space translates to the solution of $t$ Helmholtz-type equations, as given by Eqns.~\ref{Eqn:PreconditionerApplication} and \ref{Eqn:Preconditioner-Helmholtz}. For complex-valued coefficients $c_j$ and $d_j$, the conjugate of the coefficients also occur in the expansion, whereby only one of the corresponding linear systems needs to be solved.  Given that the value of $t$ and the sparsity of the Helmholtz-type operator as well as its condition number for $d_j \neq 0$ are all independent of $N$, the real-space preconditioner can be applied with modest computational effort that scales linearly with system size. In fact, the cost associated with the solution of such linear systems constitutes a minor fraction of the total cost in real-space DFT calculations, particularly since they are amenable to scalable high performance computing. In addition, given the resemblance of the above preconditioning scheme to real-space Kerker \cite{shiihara2008real,lin2013elliptic}---setting $c_0=0$, $c_1 = 1$, $d_1 \in \mathbb{R}$, and $t=1$ in Eqns.~\ref{Eqn:PreconditionerApplication} and \ref{Eqn:Preconditioner-Helmholtz} recovers the real-space Kerker preconditoner---implementation within existing real-space codes is straightforward. Also, though the focus here is on the finite-difference discretization, the proposed approach is expected to be applicable to other real-space techniques such as the finite-element method \cite{Sterne,Phanish2010,MOTAMARRI2020106853}.

%%%%%%%%%%%%%%%%%%%%%%%%%%%%%%%%%%%%%%%%%%%%%%%%%%%%%%%%%%%%%%%%%%%%%%%%%%%%%%%%%%%%%%%%%%%%%%%%%%%%%%
\section{Results and discussion} \label{Sec:Results}
We now verify the accuracy and efficiency of the proposed real-space formulation for isotropic Fourier-space preconditioners used to accelerate the SCF iteration in Kohn-Sham DFT. As representative schemes, we consider the truncated-Kerker (t-Kerker) \cite{VASP,zhou2018applicability} and Resta \cite{resta1977thomas,zhou2018applicability} preconditioners:  
\begin{align} 
\text{\emph{t-Kerker:}} & \quad \quad \tilde{P}(|\mathbf{q}|) = \max\left( \alpha_{0},\frac{|\mathbf{q}|^2}{k_{TF}^2+|\mathbf{q}|^2}\right) \,, \label{Eqn:t-Kerker} \\
\text{\emph{Resta:}} & \quad \quad \tilde{P}(\mathbf{|\mathbf{q}}|) = \frac{\frac{q_0^2 \sin(|\mathbf{q}|R_s)}{\epsilon_0 |\mathbf{q}| R_s}+|\mathbf{q}|^2}{q_0^2+|\mathbf{q}|^2} \,, \label{Eqn:Resta}
\end{align}
where $\alpha_0$ is the threshold parameter, $k_{TF}$ is Thomas-Fermi screening length, $q_0$ is a constant related to the valence electron Fermi momentum, $\epsilon_0$ is static dielectric constant, and $R_s$ is the screening length. Here, we choose $\alpha_0 = 0.17$ and $k_{TF} = 0.529$ Bohr$^{-1}$ for t-Kerker; and $\epsilon_0 = 10$, $q_0 = 0.68$ Bohr$^{-1}$, and $R_s = 6.61$ Bohr \cite{resta1977thomas} for Resta. These preconditioners have been shown to accelerate the SCF convergence in insulating as well as inhomogeneous systems \cite{VASP,zhou2018applicability}. 

We implement the proposed approach in the latest version of SPARC \cite{ghosh2017sparc1,ghosh2017sparc2}, a real-space DFT code that is highly competitive with planewave codes in terms of both accuracy and efficiency. We consider MoS$_2$ slabs and hydrogen-passivated (111) carbon slabs as representative systems. In all calculations, we employ the GGA exchange-correlation functional \cite{perdew1996generalized}; ONCV pseudopotentials \cite{hamann2013optimized} with $14$, $6$, and $4$ valence electrons for molybdenum, sulphur, and carbon, respectively; Pulay mixing \cite{anderson1965iterative,pulay1980convergence} with relaxation parameter $\beta=0.4$ and mixing history of $7$; periodic boundary conditions with vacuum of 20 \AA \, and Brillouin zone integration using a $6\times 6 \times 1$  Monkhorst-Pack grid \cite{monkhorst1976special}; and 12th-order finite-difference approximation with mesh-size of 0.3 Bohr. Since the coefficients $c_j$ and $d_j$ are complex-valued for the current parameter choices, only $t/2$ Helmholtz-type linear systems (Eqn.~\ref{Eqn:Preconditioner-Helmholtz}) need to be solved, for which we employ the AAR method \cite{pratapa2016anderson,suryanarayana2019alternating} with a tolerance of $0.01$ on the relative residual.\footnote{Tighter tolerances do not provide any noticeable gains in terms of SCF convergence.}

First, we check the quality of fit between the aforementioned isotropic Fourier-space preconditioners and their rational function approximations. Specifically, we compare the $t=2$ and $t=4$ approximations with the exact form given in Eqns.~\ref{Eqn:t-Kerker} and \ref{Eqn:Resta}. We observe that there is very good agreement between the fits and the original functions, with the curves being practically indistinguishable. Specifically, the maximum difference between the $t=2$ rational fit and the exact t-Kerker and Resta preconditioner functions is 0.04 and 0.03, respectively.  This reduces to 0.01 in both cases for $t=4$. It is therefore clear that even a low degree rational function approximation can provide a very good fit for such preconditioning functions. Indeed, we have verified that this conclusion is not just restricted to the chosen set of parameters, but is also valid for other parameter choices. 

\begin{figure}[h!]
\centering
\subfloat[t-Kerker]{\label{Fig:mod:ker}\includegraphics[keepaspectratio=true,width=0.43\textwidth]{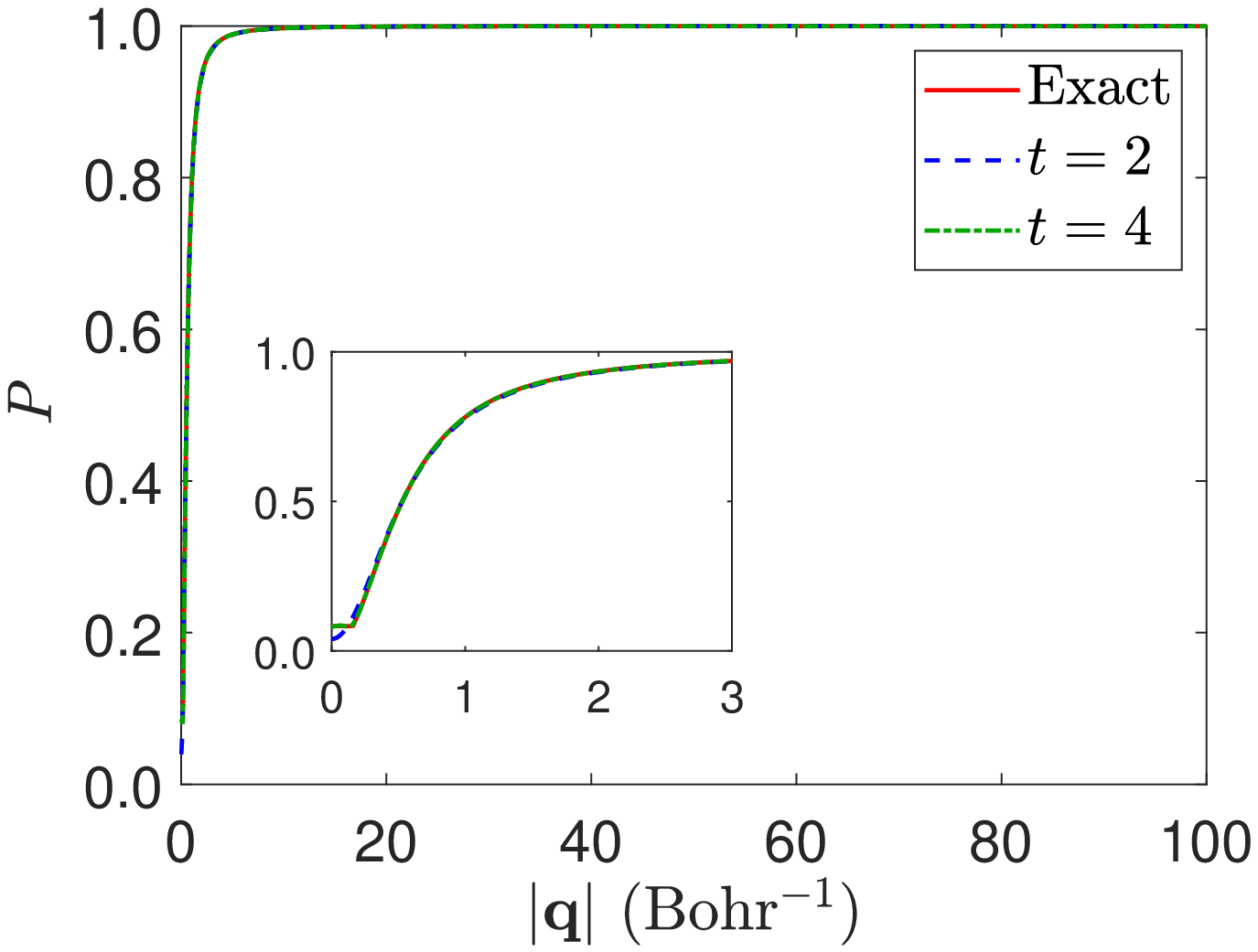}} \hspace{0.05\textwidth}
\subfloat[Resta]{\label{Fig:Resta}\includegraphics[keepaspectratio=true,width=0.43\textwidth]{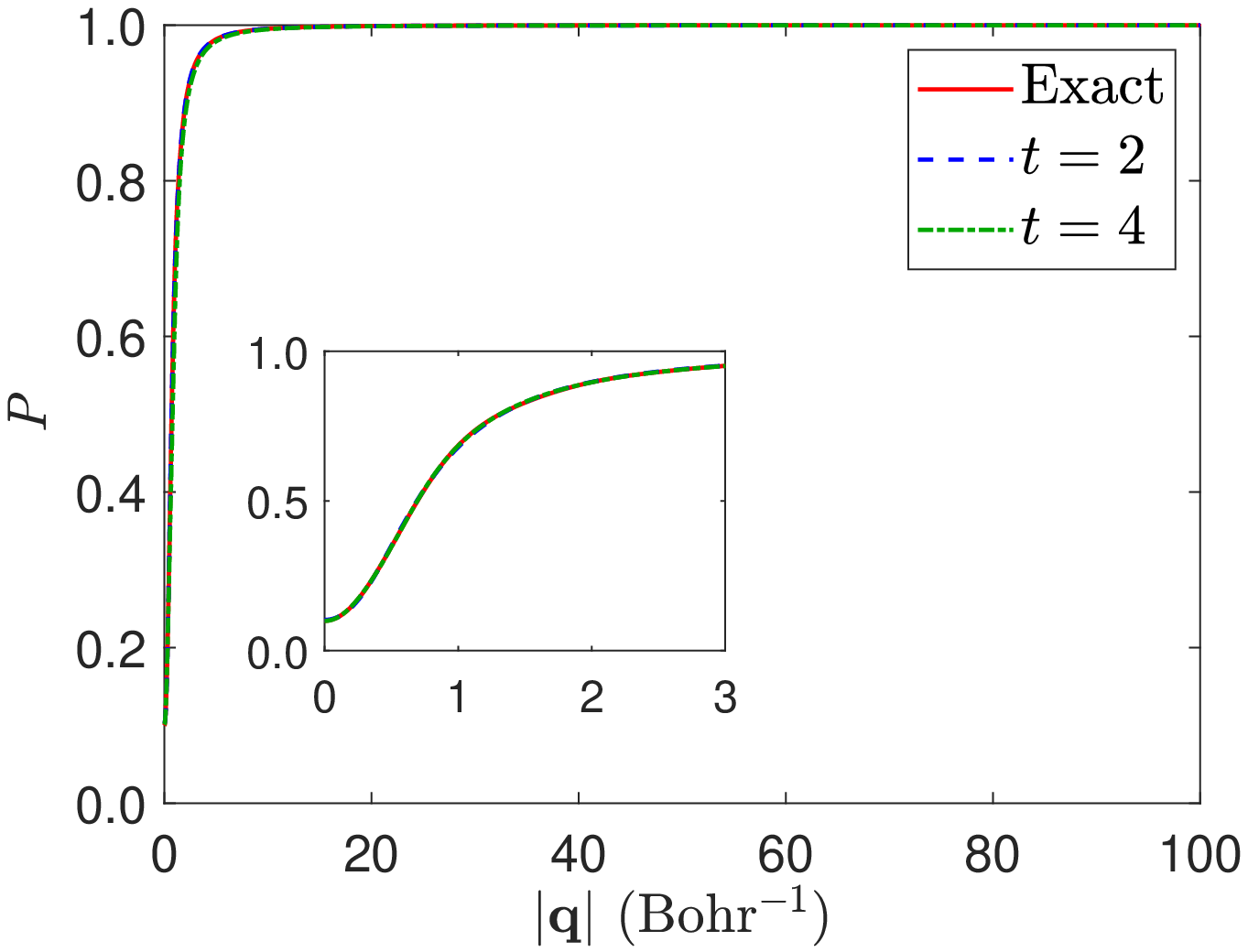}}
\caption{The exact preconditioner function and its rational function approximation. }
\label{Fig:ker:Res}
\end{figure}

\begin{figure}[ht!]
\centering
\subfloat[5-layer MoS$_2$ slab, Resta] 
{\label{Fig:MoS2realfourier}\includegraphics[keepaspectratio=true,width=0.43\textwidth]{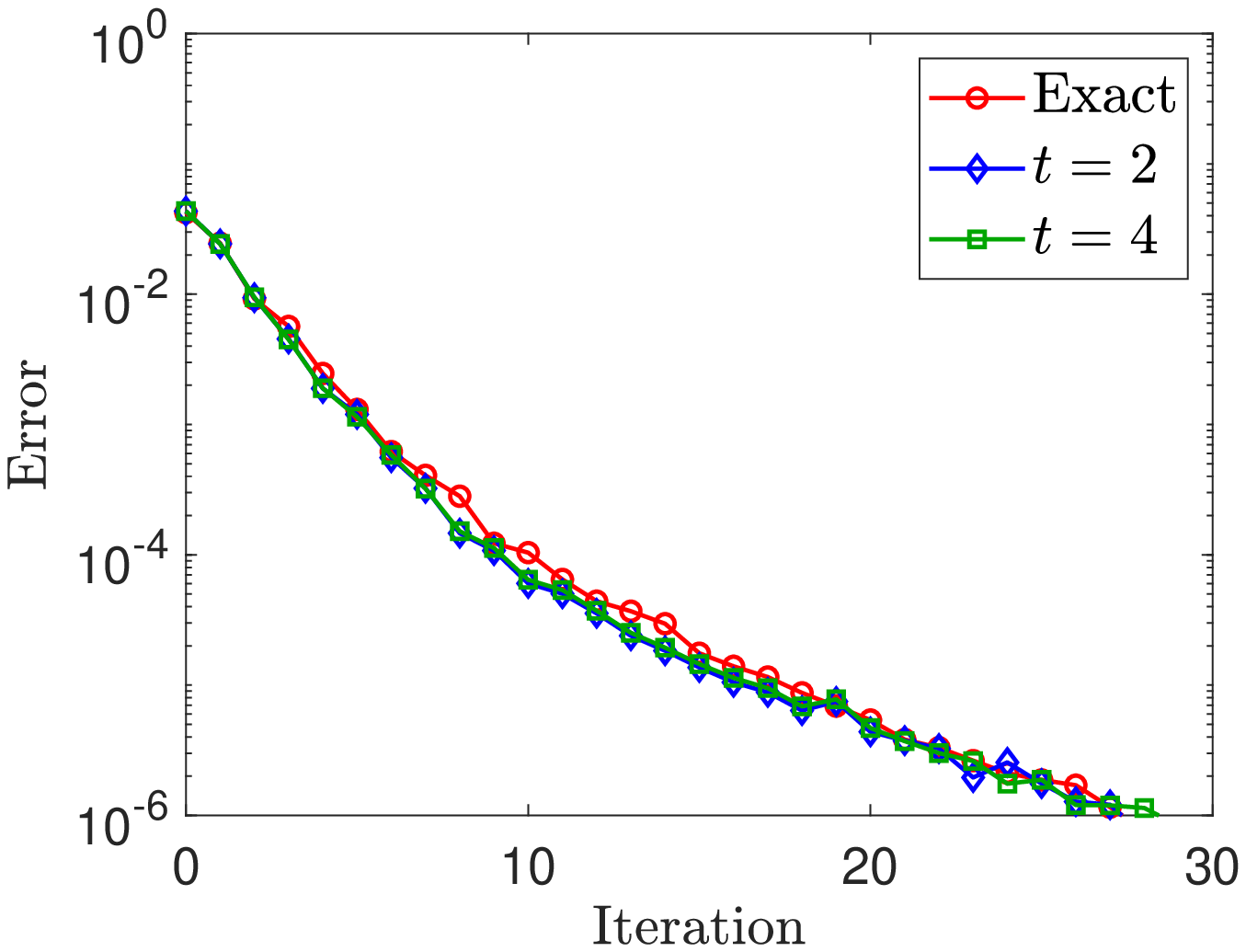}} \hspace{0.05\textwidth}
\subfloat[20-layer carbon slab, t-Kerker]{\label{Fig:Sirelafourier small}\includegraphics[keepaspectratio=true,width=0.43\textwidth]{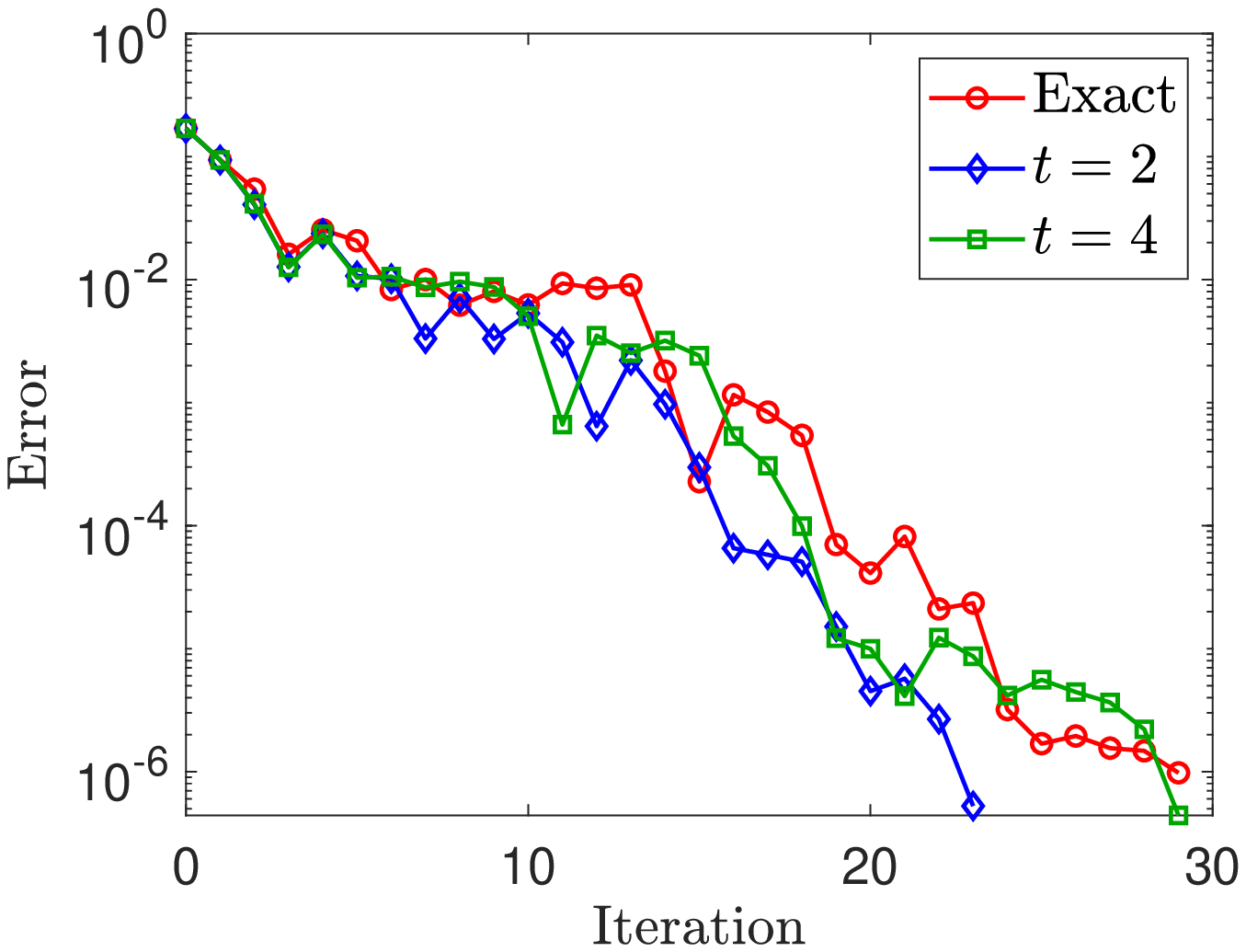}}
\caption{Comparison of the SCF error as a function of iteration number for the exact Fourier-space preconditioner and its real-space formulation. The SCF error is defined to be the normalized residual in the electron density.}
\label{Fig:comparisonrealfourier}
\end{figure}

Next, we establish that the approximation involved in the rational function fit does not negatively impact the convergence of the SCF iteration in DFT calculations. For this study, we choose a 5-layer MoS$_2$ slab (30 atoms) and a 20-layer hydrogen-passivated carbon slab (44 atoms) as representative examples. In Fig.~\ref{Fig:comparisonrealfourier}, we present the progression of the error during the SCF iteration for exact Fourier-space preconditioning and compare it with the real-space formulation corresponding to rational function approximations of $t=2$ and $t=4$. We perform exact preconditioning in Fourier-space, i.e., we take the Fourier transform of $\langle\mathbf{f}_{k}\rangle$, apply the preconditioning in Fourier-space, and then take the inverse Fourier transform. It is clear from the progression of the SCF iteration in Fig.~\ref{Fig:comparisonrealfourier} that the proposed method can acclerate the SCF iteration to the same extent as the exact Fourier-space analogue, further verifying the accuracy of the developed approach.  

\begin{figure}[h!]
\centering
\subfloat[13-layer MoS$_2$ slab]{\label{Fig:MoS2scfcomparison}\includegraphics[keepaspectratio=true,width=0.43\textwidth]{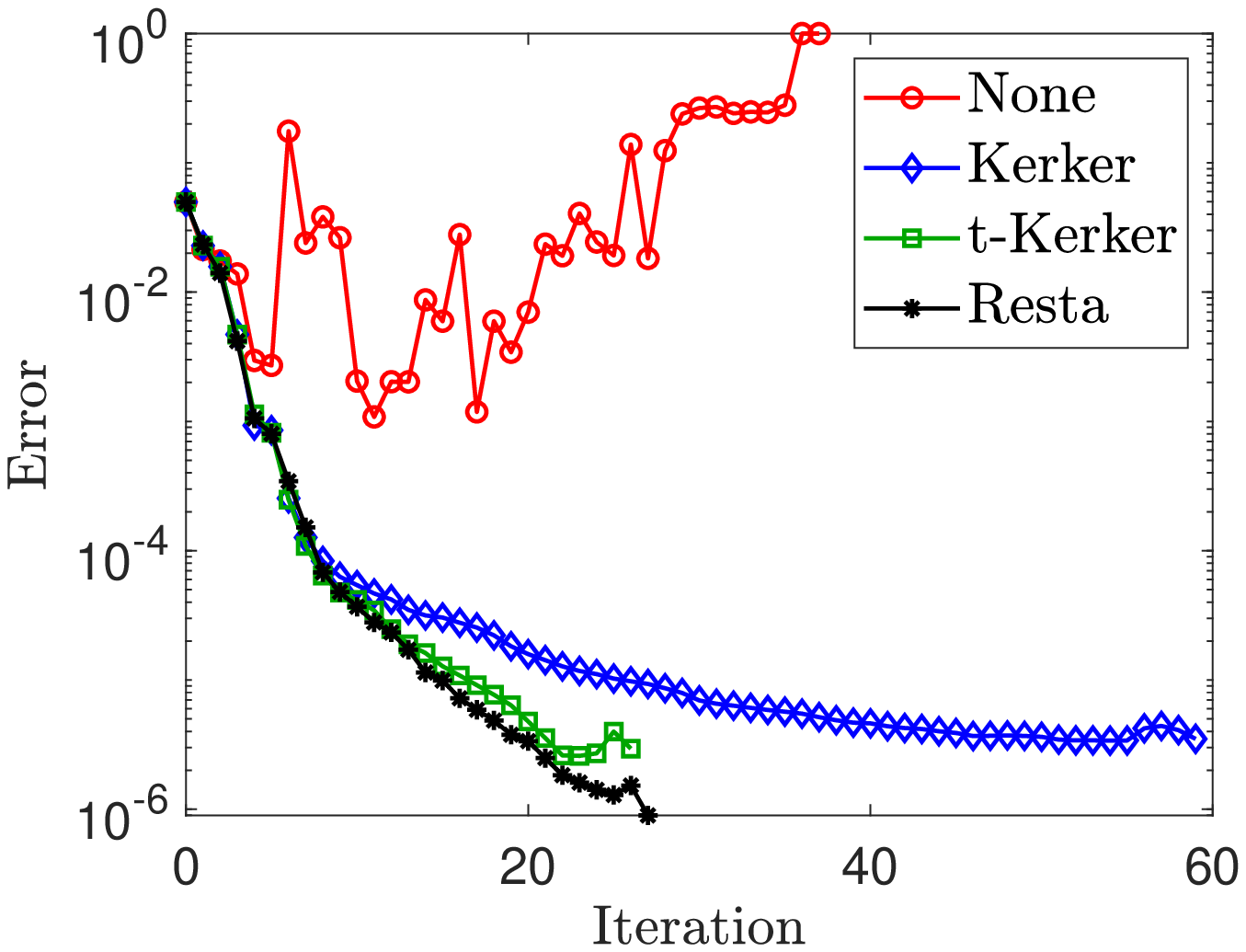}} \hspace{0.05\textwidth}
\subfloat[30-layer carbon slab]{\label{Fig:Siscfcomparison}\includegraphics[keepaspectratio=true,width=0.43\textwidth]{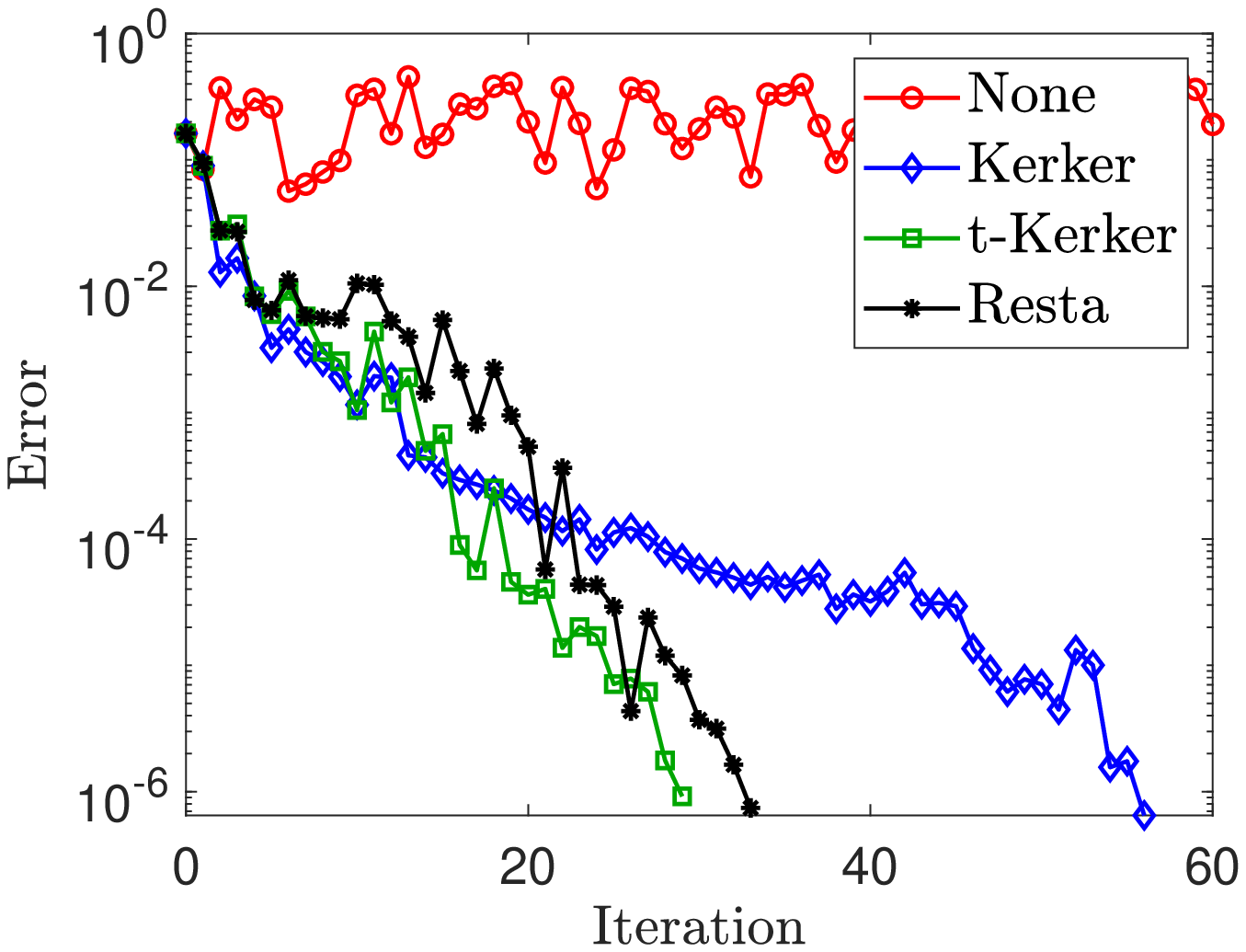}}
\caption{Comparison of the SCF error as a function of iteration number for the following cases: no preconditioner, and real-space Kerker, t-Kerker, and Resta preconditioners. The SCF error is defined to be the normalized residual in the electron density. } 
\label{Fig:scfcomparison}
\end{figure}

Next, we show that the proposed approach can be used to accelerate the SCF convergence in real-space DFT calculations. For this study, we choose a 13-layer MoS$_2$ slab (78 atoms) and a 30-layer hydrogen-passivated carbon  slab (64 atoms). We use a rational function approximation corresponding to $t=2$.  In Fig.~\ref{Fig:scfcomparison}, we present the convergence of the SCF iteration for the following cases: no preconditioner, and real-space Kerker \cite{shiihara2008real,lin2013elliptic}, t-Kerker, and Resta preconditioners. We observe that the t-Kerker and Resta preconditioners start outperforming Kerker as the SCF error reduces, demonstrating their utility. This behavior can be attributed to the difference between the preconditioners in the low frequency regime, errors of which type are the last to be removed from within the SCF method. Note that the Kerker preconditioner effectively reduces to t-Kerker for small to moderate sized systems, which explains its effectiveness for insulating systems, like the ones studied here. Though they are insulating, convergence is not achieved in the absence of a preconditioner, likely due to the inhomogeneous nature of the systems. Indeed, the mixing parameter can be reduced to enable convergence, e.g.,  the SCF converges to an error of $10^{-4}$ in $87$ iterations for the carbon system when $\beta=0.2$.  

\begin{figure}[h!]
\centering
\subfloat[MoS$_2$ slab, Resta]{\label{Fig:MoS2scaling}\includegraphics[keepaspectratio=true,width=0.43\textwidth]{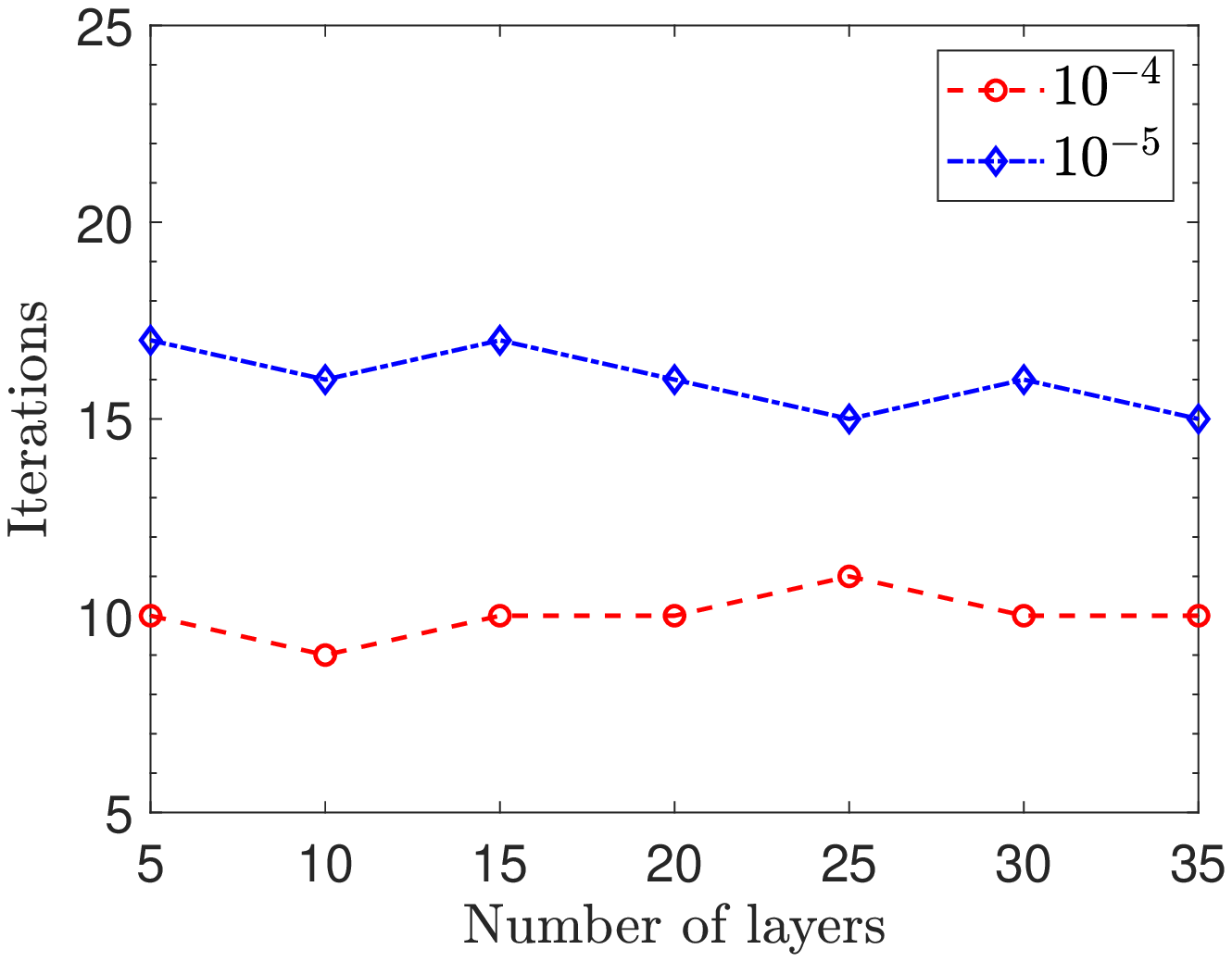}} \hspace{0.05\textwidth}
\subfloat[Carbon slab, t-Kerker]{\label{Fig:Cscaling}\includegraphics[keepaspectratio=true,width=0.43\textwidth]{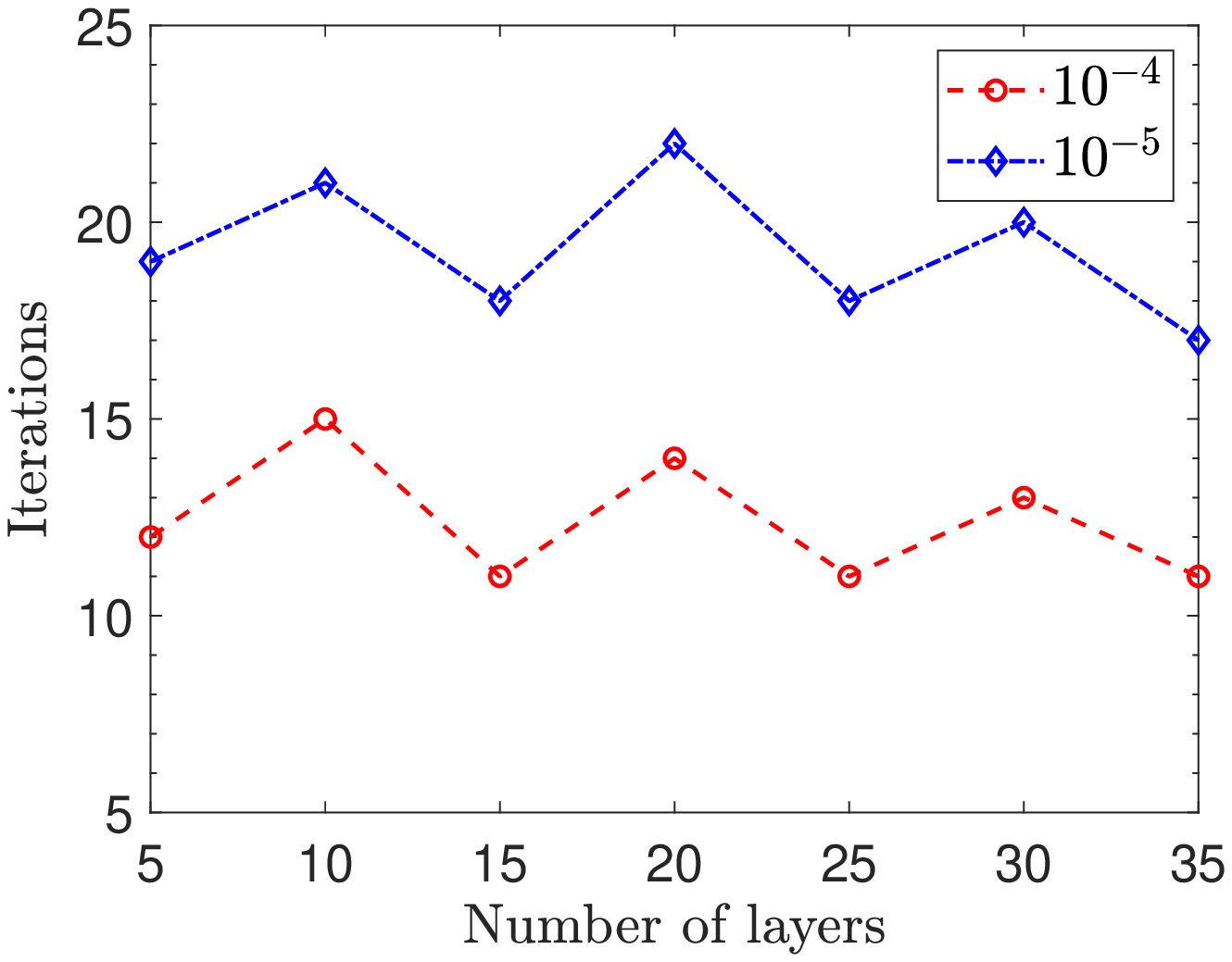}}
\caption{Variation in the number of SCF iterations as a function of system size for errors of $10^{-4}$ and $10^{-5}$. The SCF error is defined to be the normalized residual in the electron density. } 
\label{Fig:scfiterationscaling}
\end{figure}

Finally, we verify the effectiveness and efficiency of real-space preconditioning. To do so, we consider MoS$_2$ and hydrogen-passivated carbon slabs ranging from $5$ to $35$ layers, and determine the number of SCF iterations required for convergence. It is clear from the results in Fig.~\ref{Fig:scfiterationscaling} that the number of iterations are essentially independent of system size, demonstrating the effectiveness of the real-space preconditioner. Its efficiency relative to the exact Fourier-space analogue is dictated by the number of linear solver iterations and processors used for the DFT calculation. Since the condition number of the Helmholtz-type operator (Eqn.~\ref{Eqn:Preconditioner-Helmholtz}) is typically small and independent of system size; the tolerance used for the linear solver does not have to be strict; and moderate to large-scale parallelization is commonly employed for  DFT calculations, the real-space formulation is appealing. For the examples considered here, the condition number is $\sim 650$ and the number of iterations required by AAR (no preconditioner) is $\sim 35$. The crossover between the real- and Fourier-space formulations occurs at hundreds of processors, e.g., the real-space approach is $\sim 1.25 \times$ faster for the 5-layer MoS$_2$ system on $600$ processors when the data reorganization time for FFT is included, with larger gains as the number of processors is increased. Overall, the proposed real-space approach is attractive for large-scale DFT calculations, where correspondingly large parallelization is naturally required.

%%%%%%%%%%%%%%%%%%%%%%%%%%%%%%%%%%%%%%%%%%%%%%%%%%%%%%%%%%%%%%%%%%%%%%%%%%%%%%%%%
\section{Concluding remarks} \label{Sec:ConcludingRemarks}
In this work, we have presented a real-space formulation for isotropic Fourier-space preconditioners used to accelerate the self-consistent field iteration in DFT simulations. Specifically, we have approximated the preconditioner in Fourier space using a rational function, and have then expressed its real-space application in terms of the solution of sparse Helmholtz-type linear systems of equations. Using the truncated-Kerker and Resta preconditioners as representative examples, we have demonstrated that the proposed real-space method is both accurate and efficient. In particular, we have shown that the application of these preconditioners in real-space requires the solution of only one linear system, while accelerating self-consistency to the same degree as their exact Fourier-space counterparts. Overall, this paper overcomes one of the limitations of real-space approaches, further increasing their attractiveness for DFT calculations.

%%%%%%%%%%%%%%%%%%%%%%%%%%%%%%%%%%%%%%%%%%%%%%%%%%%%%%%%%%%%%%%%%%%%%%%%%%%%%%%%%%
\section*{Acknowledgement}
The authors gratefully acknowledge the support provided by the National Science Foundation (NSF) under grant number 1663244.
%%%%%%%%%%%%%%%%%%%%%%%%%%%%%%%%%%%%%%%%%%%%%%%%%%%%%%%%%%%%%%%%%%%%%%%%%%%%%%%%%%

\bibliographystyle{elsarticle-num}
\bibliography{Manuscript}

\end{document}